\documentclass[11pt]{article}
\pdfoutput=1
\usepackage{jcappub,cancel}
\usepackage{paralist}
\usepackage{booktabs}
\usepackage{aas_macros}
\usepackage[a4paper]{geometry}
\graphicspath{{images/}}
\usepackage{subfigure}
\usepackage{hyperref}

\title{Where do the 3.5 keV photons come from? A morphological study of the Galactic Center and of  Perseus}

\author[a,b]{Eric Carlson,}
\author[a,b]{Tesla Jeltema,}
\author[a,b]{Stefano Profumo}
\affiliation[a]{Department of Physics, University of California, Santa Cruz\\1156 High St, Santa Cruz, CA 95064}
\affiliation[b]{Santa Cruz Institute for Particle Physics,\\ 1156 High St, Santa Cruz, CA 95064}
\emailAdd{erccarls@ucsc.edu}
\emailAdd{tesla@ucsc.edu}
\emailAdd{profumo@ucsc.edu}

\begin{document}

\abstract{We test the origin of the 3.5 keV line photons by analyzing the morphology of the emission at that energy from the Galactic Center and from the Perseus cluster of galaxies. We employ a variety of different templates to model the continuum emission and analyze the resulting radial and azimuthal distribution of the residual emission. We then perform a pixel-by-pixel binned likelihood analysis including line emission templates and dark matter templates and assess the correlation of the 3.5 keV emission with these templates. We conclude that the radial and azimuthal distribution of the residual emission is incompatible with a dark matter origin for both the Galactic center and Perseus; the Galactic center 3.5 keV line photons trace the morphology of lines at comparable energy, while the Perseus 3.5 keV photons are highly correlated with the cluster's cool core, and exhibit a morphology incompatible with dark matter decay. The template analysis additionally allows us to set the most stringent constraints to date on lines in the 3.5 keV range from dark matter decay.}

\maketitle

\section{Introduction}
Earlier this year, an unidentified spectral line at 3.55-3.57 keV was reported by Bulbul et al~\cite{Bulbul:2014sua} using {\em XMM-Newton} observations of galaxy clusters.  The signal was detected in five independent samples, including observations both of single objects and stacked cluster spectra.  Subsequently, the presence of an unidentified line at $\sim3.52$ keV was confirmed by a re-analysis of {\em XMM} data of the Perseus cluster, and also reported from an analysis of {\em XMM} observations of M31 in Ref.~\cite{Boyarsky:2014jta}. {\em XMM} blank-sky observations do not reveal any evidence for any feature at the energy of interest \cite{Boyarsky:2014jta}.  Furthermore, when detected in clusters, the line is observed to have a redshift consistent with that of the host cluster, helping to rule out an obvious instrumental origin~\cite{Bulbul:2014sua}.

Two possible explanations have dominated the ensuing debate over the origin of this mysterious signal.  The first was initially put forth by Ref.~\cite{Bulbul:2014sua}, proposing the exciting possibility that the line is due to the decay of sterile neutrino dark matter.  The second possibility is astrophysical, suggesting that the size and impact of uncertainties  in modeling the complex multi-phase plasma structure in clusters had been previously underestimated, and that one or more weak plasma transition lines could in fact be  responsible for the signal~\cite{bananas}.  Below we summarize these ongoing developments and present a novel spatial analysis method which aims to discriminate the two scenarios using existing archival {\em XMM} X-ray data.

Self-consistency of the unidentified X-ray line across different objects and for different X-ray instruments ({\em XMM} MOS, PN and Chandra) has appeared problematic from the start. For example, the precise spectral location of the line\footnote{In what follows, we will refer to the energy of the line simply as 3.5 keV for brevity.} has been difficult to pin down and the line flux for a given object were found to vary substantially even between different instruments on the same telescope.  Interpretations of the signal in terms of a standard decaying dark matter candidate are also troubling due to the observed strength being inconsistent across different objects~\cite{Bulbul:2014sua}.

The possibility of explaining the 3.5 keV signal with new physics prompted vigorous activity especially in the model-building community, and a variety of possible particle physics scenarios have been proposed. In particular, models with sterile neutrinos can potentially provide avenues to successful baryogenesis via leptogenesis, to the generation of the observed pattern of (active) neutrino masses and mixing, and to providing a dark matter candidate which can alleviate certain small-scale issues of cold dark matter scenarios \cite{Boyarsky:2009ix}. 
Model building efforts were also directed toward constructing a consistent new-physics picture that could reconcile varying signal strengths across different objects, especially in the context of axion-like particles converting to 3.5 keV photons in the presence of magnetic fields (see e.g. Ref.~\cite{Higaki:2014zua, Jaeckel:2014qea, Lee:2014xua, Cicoli:2014bfa, Ringwald:2014vqa} for early studies of this scenario).

In the dark-matter decay picture the relative signal strength is easily calculated, scaling as the dark matter density integrated over the line of sight. It thus became clear that a confirmation of the non-standard and possibly dark-matter-related origin of the unidentified line might come from observations of nearby targets such as the Galactic center (GC), or nearby dark-matter-dominated systems, such as local dwarf spheroidal galaxies (dSph), where no plasma emission line background is expected.

Ref.~\cite{Riemer-Sorensen:2014yda} proceeded to analyze Chandra data of the Galactic center, finding no evidence for an excess 3.5 keV line after including known plasma emission lines. This study produced constraints on a decaying dark matter scenario in clear tension with such an interpretation of previous results, but it was not clear from this work if the measured line fluxes near 3.5 keV were consistent with those expected from plasma lines. 

While relevant, Chandra observations of the Galactic center are significantly shallower than available archival {\em XMM} observations. In Ref.~\cite{bananas}, two of us (TJ \& SP) analyzed {\em XMM} observations of the center of the Galaxy, and discovered a line at an energy of about 3.5 keV. We pointed out that the detected line signal was compatible, at face value, with a dark matter decay origin, and calculated the corresponding preferred sterile neutrino lifetime and mixing angle\footnote{This aspect of our study appears to have been under-appreciated in the recent literature on the topic.}. However, we also showed that, given the fluxes of other bright plasma emission lines, the flux of the detected 3.5 keV line is compatible with two atomic lines from K XVIII at 3.47 and 3.51 keV, making astrophysical plasma emission a natural explanation of the 3.5 keV feature.

In Ref.~\cite{bananas} we also re-analyzed, in the 3-4 keV energy range, archival {\em XMM} data from M31, and found no statistical evidence for a line at 3.5 keV. In addition, we pointed out that the procedure employed in Ref.~\cite{Bulbul:2014sua} to predict the K XVIII line flux and the flux of other atomic transition lines was not, contrary to what stated in that paper, maximally conservative. We argued that the multi-temperature models employed there were biased towards excessively large temperatures which artificially suppress, for example, the expected K XVIII line intensity\footnote{We further demonstrate this point in Ref.~\cite{tjspreply}, where we show that the Ca XX to Ca XIX ratio indicates temperatures inconsistent with the multi-temperature models employed in Ref.~\cite{Bulbul:2014sua}: for example, in the case of the ``all-clusters'' MOS sample, the large temperatures employed in Ref.~\cite{Bulbul:2014sua} produce {\em under-}estimates of the K XVIII line flux by factors between more than 4 and more than 10.}.

Two key subsequent observational analyses of other sources also found no confirmation of an exotic origin for the 3.5 keV line (or any line at that energy whatsoever): Ref.~\cite{Malyshev:2014xqa} analyzed stacked observations of dwarf galaxies, while Ref.~\cite{Anderson:2014tza} analyzed Chandra and {\em XMM} observations of two large samples of galaxies and groups of galaxies, which should exhibit tenuous, if any, plasma emission features. Both results robustly rule out a dark matter decay interpretation of the 3.5 keV line observations reported by Ref.~\cite{Bulbul:2014sua} and by Ref.~\cite{Boyarsky:2014jta}. 

Finally, a recent study utilized deep Suzaku observations of Perseus, Coma, Virgo and Ophiuchus, to test the dark matter decay hypothesis for the 3.5 keV line \citep{Urban:2014yda}. While confirming the existence of a 3.5 keV signal in Perseus, Ref.~\cite{Urban:2014yda} did not find any evidence of an associated signal, with the appropriately rescaled intensity, in the other three clusters, robustly ruling out a dark matter interpretation for the 3.5 keV line observed from Perseus; also, Ref.~\cite{Urban:2014yda} points out that  the radial variation of the 3.5 keV signal in Perseus is in tension with what expected from dark matter decay, and finds evidence that the Perseus signal could be potentially explained by elemental lines, as we had originally suggested in Ref.~\cite{bananas}.

Following our analysis of the GC data in Ref.~\cite{bananas}, two comments, Ref.~ \cite{Boyarsky:2014paa, Bulbul:2014ala}, appeared regarding our paper, as well as an analysis of {\em XMM} data of the Galactic center, which confirmed our discovery of a 3.5 keV line \cite{BoyarskyGC}.  While these comments debate some details of the analysis in Ref.~\cite{bananas}, we show in Ref.~\cite{tjspreply} that none of our conclusions are substantially affected or challenged by the points raised.  This ongoing discussion highlights the difficulty in assessing the origin of (and in the case M31, the existence of) such a weak spectral feature amongst a background rich in astrophysical emission. 

Previous studies have argued that a final word on the origin of the 3.5 keV line might possibly come from future high-resolution instruments such as Astro-H.  In this work, we show that it is possible to assess the nature of the 3.5 keV line by using available archival data in a completely independent context compared with the spectral analyses performed thus far: we show that a {\em morphological} analysis of the 3.5 keV emission allows for a critical evaluation of the physical nature of the source, and conclude that the 3.5 keV line is highly unlikely to originate from dark matter decay, or from axion-like particle conversion.  We also derive the most constraining limits to-date on the lifetime of a putative radiatively decaying dark matter particle producing a monochromatic 3.5 keV line.

In this study we carry out an extensive morphological analysis of two of the most tantalizing locations where the 3.5 keV line has been conclusively detected: the Galactic center and Perseus cluster. As the 3.5 keV line is quite tenuous compared to the continuum emission, the choice and determination of the continuum model is important. We thus consider a broad  sample of continuum emission models, and assess possible systematic effects associated with modeling the morphology of the continuum using different assumptions. We then study the morphology of plasma emission lines observed at different energies, and assess the radial and azimuthal behavior of the  residual emission at 3.5 keV. We compare this residual emission with the morphology of line emission templates as well as with dark matter templates. Finally, we study the cross-correlation of the 3.5 keV morphology with a sliding-window template, and calculate the resulting constraints on any emission sharing the morphology expected from dark matter decay models for both the Galactic center and Perseus.


\section{Methodology}

In this section, we introduce the two targets we focus on in the present study, we describe our {\em XMM} data selection, the details of the binned-likelihood analysis, the  ``sideband'' templates for nearby spectral lines and continuum emission, and finally, the generation of sky-maps for decaying dark matter based on several choices for the dark matter halo model.  

\subsection{Choice of Targets}

The present study focuses on the Galactic center \cite{bananas}, and on the Perseus cluster \cite{Bulbul:2014sua, Boyarsky:2014jta}. The proximity of the Galactic center and the inherent asymmetry of the expected astrophysical emission makes it an especially compelling target for a morphological study; the Perseus cluster's 3.5 keV line has been robustly detected with several X-ray instruments (including {\em XMM} MOS but not PN, Chandra ACIS-S and ACIS-I, and Suzaku) and the cluster has a large enough extension in the sky, compared to the instrumental angular resolution, that a morphological study is also possible.

If the 3.5 keV line indeed arises from the decay (or annihilation, or conversion of the decay products) of particle dark matter, the spatial morphology of any excess emission must be both roughly spherical\footnote{A possible exception, which we discuss in what follows, is the case of the Galactic center for axion-like-particle conversion.} and declining with growing distance from the center of the halo. On the other hand, the flux from atomic emission lines should trace the distribution of the astrophysical plasma as the product of atomic abundance and emissivity, the latter of which possesses a strong dependence on the local plasma temperature.

The X-ray emission from the Galactic center is dominated by resolved and unresolved X-ray point sources as well as thermal emission from the hot gas.  In addition to hot plasma and high gas densities, the Galactic center region is densely packed with feedback sources, including the central black-hole, Sgr A*, supernova remnants such as Sgr A East, and stellar clusters.  This environment leads to an extremely complex and spatially varying multi-temperature plasma structure, as well as to strong gradients in the atomic abundances of each radiating species, making it difficult to conclusively spatially correlate strong emission lines~\cite{sgrA:1,sgrA:2,sgrA:3}.  

X-ray emission from galaxy clusters originates from the hot intergalactic medium heated during gravitational collapse.  With very different cooling and feedback processes compared to the Galactic center, such systems provide an independent lens for spectrally {\em and} morphologically differentiating dark matter emission from that of $\sim$keV plasma emission lines. Already, there is substantial tension between the 3.5 keV line fluxes detected within the stacked and unstacked observations of clusters~\cite{Bulbul:2014sua}, perhaps providing an indication of differing multi-phase plasmas, or more exotic dark matter candidates, such as an Axion-Like-Particle with a photo-conversion rate dependent on the cluster's magnetic field structure.

Categorically, clusters fall into two groups. `Cool-core clusters' are stable virialized structures with no recent major-mergers, whereas in `non-cool-core' clusters recent collisions have inhibited dynamical relaxation.  Radiative cooling timescales are $\sim10^8-10^9$~yr, creating a flow of hot ($\sim7$~keV) dispersed gas to into the cool ($\sim3$~keV) cluster center.  AGN feedback eventually halts the inflow and maintains temperatures of about a few keV~\cite{Fabian:2003}.  Perseus is a prototypical cool-core cluster with a complicated core structure, where the plasma has clearly been affected by feedback from the bright central AGN, offering an excellent test case for a comparative study of DM versus gas-correlated emission.  Unfortunately, due to the relatively low flux of the spectral line in question, all but the most prominent of the detailed core structures in Perseus are unresolved, making a morphological analysis potentially less stringent than in the Galactic center. A prominent spatial feature associated with the cluster core would, however, provide a strong handle for differentiating exotic emission from that associated with the cool core~\cite{Fabian:2000,Fabian:2002,Fabian:2003} (see also Ref.~\cite{Urban:2014yda}).

\subsection{X-Ray Data Selection}
\label{subsec:xmm}

We employ archival observations from {\em XMM} for both the Galactic center and the Perseus cluster.  We restrict our analysis to using the two EPIC MOS detectors which have both higher spatial resolution and fewer instrumental chip gap features than the EPIC PN detector.  While in principle the Chandra Observatory would offer better spatial resolution, the smaller field-of-view and lower effective area of Chandra make {\em XMM} the better choice.  In addition, given the small 3.5 keV line flux, we employ a binning that is in any case larger than the inherent instrumental angular resolution.

For the Galactic center we use the same observations as listed in Table 1 of Ref.~\cite{bananas}.  This set has been chosen to eliminate observations with either strong particle background flaring or significant flaring from Sgr A*.  For the Perseus cluster we utilize the same two {\em XMM} observations as in Ref.~\cite{Bulbul:2014sua}.  The basic data reduction follows Ref.~\cite{bananas}.  All observations were reprocessed using the {\tt emchain} task in {\em XMM} SAS version 13.5.0.  Particle background flares were then removed using the {\tt mos-filter} task in the {\em XMM} {\tt ESAS} package \citep{esas,esas2}.  The energies of all photons in the Perseus field-of-view are then blue-shifted by the current NED value\footnote{\href{http://ned.ipac.caltech.edu}{http://ned.ipac.caltech.edu}} $z=0.0179$.

For each observation, we also created a mask to remove regions containing bright point sources, including Sgr A* in the Galactic center and the central AGN in Perseus, as well as low exposure regions due to chip gaps and bad columns on the detector.  Masks were generated with the {\tt ESAS} task {\tt cheese} run on broad-band (0.4-7.2 keV) images.  These masks are applied consistently to all images generated as described below.  To appropriately include the detector response when modeling the possible dark matter contribution (see Sec.~\ref{subsec:dm_templates}), we also created exposure maps in the 3-4 keV band for each observation.

\subsection{Binned Likelihood Analysis}

In order to quantify the contributions of various templates to the 3.45-3.60 keV band, we perform a pixel-by-pixel binned likelihood analysis.  Photons are binned into 20'' square pixels over the $\approx0.5^\circ$ diameter field of view of {\em XMM}'s MOS1 \& MOS2 imagers.  This choice provides an average of several photons per pixel for the weak detectable emission lines (after subtracting off continuum emission), and happens to be only slightly larger than the {\em XMM} MOS point-spread function, implying that little would be gained with finer resolutions.  We verified that the main conclusions reached by the present analysis are largely independent of this specific choice of binning, with only minor changes to e.g. limits.  In these cases, we have ensured that the results presented here reflect the most conservative values.

The Poisson likelihood function is defined by 
\begin{equation}
\mathcal{L}=\prod\limits_{i=1}^{N_{\rm pix}}\frac{\mu_i^{d_i} e^{-\mu_i}}{d_i!}, 
\end{equation}
where ${\mu_i}$ and $d_i$ are the predicted and observed counts for pixel $i$.  The model counts are derived by summing the included templates with independent normalizations. The normalizations of each template are restricted to be positive semi-definite and are varied until the log-likelihood is maximized.  To obtain a fast approximate fit, we use the well tested minimization algorithm {\tt MIGRAD} from the {\tt IMINUIT} python package\footnote{\href{http://iminuit.github.io/iminuit/index.html}{http://iminuit.github.io/iminuit/index.html}}.  Due to the large fit parameter space and the potential for nearly degenerate templates, we then use the stochastic global minimizer {\tt BasinHopping-L-BFGS-B}\footnote{\href{http://docs.scipy.org/doc/scipy-dev/reference/generated/scipy.optimize.basinhopping.html}{http://docs.scipy.org/doc/scipy-dev/reference/generated/scipy.optimize.basinhopping.html}} for further refinement.

A test-statistic (TS) is defined through the likelihood-ratio test TS$=-2\Delta \ln(\mathcal{L})$ where $\Delta \ln(\mathcal{L})$ is the difference in log-likelihood between the null and trial model.  When only considering the normalization of a single template, the TS behaves as a $\chi^2/2$ distribution with one degree of freedom.  Limits are found by increasing the normalization of the (e.g. dark matter) template until the log-likelihood is decreased by $s^2/2$, for an $s$ standard deviation result.  We use $s=2$ throughout, corresponding to 95\% confidence levels.

\subsection{Continuum Templates}
\label{sec:continuum}
We select 5 continuum emission bands at energies between 3.19 and 4.8 keV\footnote{See also Fig.~\ref{fig:decomp}, where we show the bands we employ as vertical green bands above 3 keV.}.  The lower-energy sideband includes photons from 3.19 and 3.27 keV.   Although partially overlapping with the bright Ar XVII line at 3.14 keV, this band is included to provide continuum photons at low energies, where dense line structure prohibits a single band of wide energy (and also prohibits the selection of line-free continuum at energies below 3 keV).  The second low-energy continuum band is also selected to be very narrow, ranging from 3.373-3.45 keV, but provides a reasonable background sample immediately neighboring our spectral region of interest. For the next neighboring continuum band, we choose 3.6-3.811~keV which also includes a series of three weak and tightly-packed Ar XVII lines.  The limited energy resolution of the {\em XMM} MOS sensors make it difficult to separate such lines, and we include them as a single wide-band contribution.  Finally, two additional high-energy bands from 4.2-4.5~keV and 4.5-4.8 keV provide a region which is effectively free of emission lines.  

\begin{figure}    \subfigure[Galactic center]{\includegraphics[width=.5\textwidth]{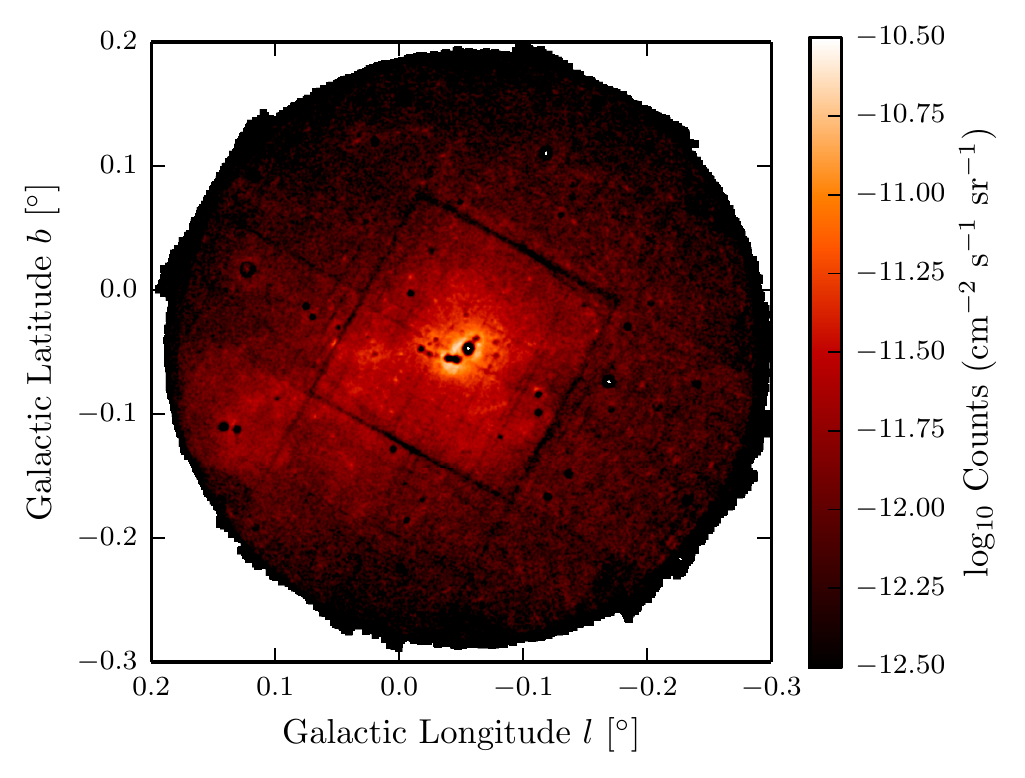}}
    \subfigure[Perseus cluster]{\includegraphics[width=.5\textwidth]{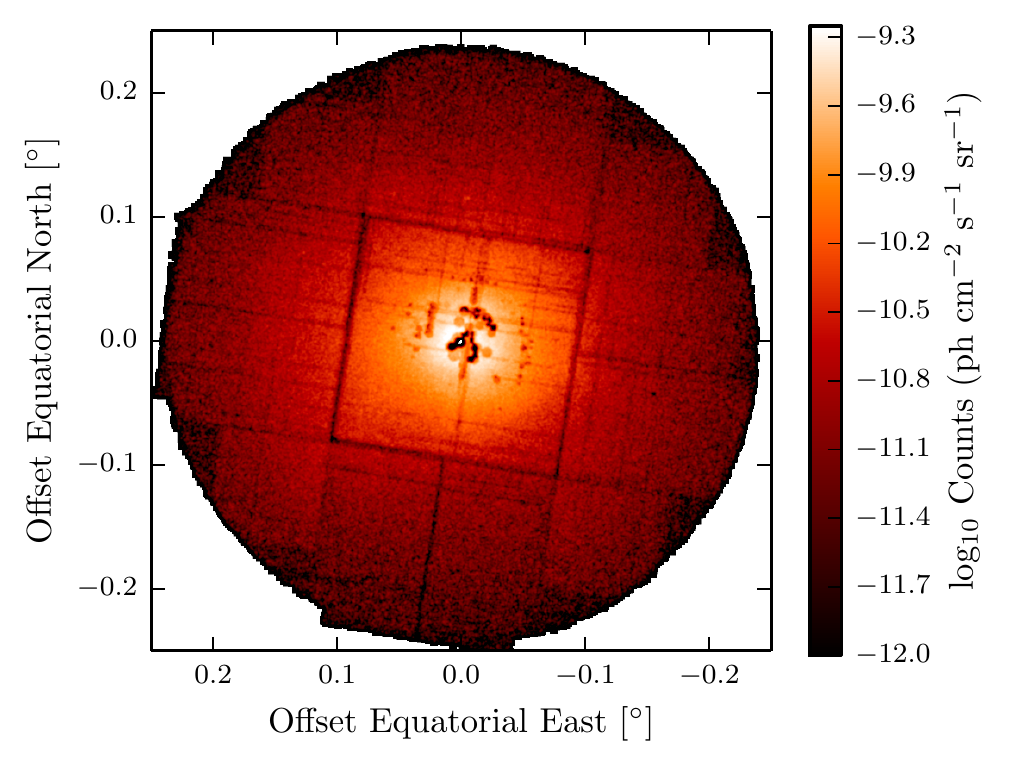}}
    \caption{Representative continuum photon maps for energies in the 4.2-5.5~keV range,  centered on Sgr A* in the Galactic center (left) region and on the Perseus cluster (right).  Count maps have been blurred using a Gaussian kernel with $\sigma=2.5''$ and are log-scaled to highlight distinctive spatial features.  In the right panel, the x-y axes are aligned with the equatorial axes and show the angular displacement from the cluster center. In both cases, bright point sources have been masked out.}
    \label{fig:continuum_sample}
 \end{figure}

While the continuum templates are visually difficult to distinguish, it is clear from the likelihood analysis that they do supply linearly independent degrees of freedom, and that the inclusion of each additional continuum template improves the model fit to photons in the 3.5 keV band at high-significance.  In Figure~\ref{fig:continuum_sample} we show log-scaled continuum count maps at high binning resolution (2.5'') for the photons between 4.2 and 5.5 keV in both the Galactic center and Perseus fields.  The GC is reveled to be very asymmetrically-structured, while the Perseus cluster is significantly smoother and more spherically symmetric (i.e. virialized).  A similar trend is observed in the morphology of emission lines, although in the Perseus cluster different plasma lines show different characteristic radii depending of the temperature at which their emissivity peaks. 

When analyzing the morphology of the residual emission (i.e. the emission not associated with continuum templates), when performing dark matter fits, and when computing limits, we consider three different continuum models with the intent of probing the sensitivity to the background model. The set of continuum models we employ is as follows:
\begin{itemize}
\item `{\em All}', utilizes all five continuum templates
\item `{\em Neighboring}' uses the two templates bracketing our 3.45-3.6 keV band of interest plus a single high-energy band (4.2-4.5 keV), and 
\item `{\em High-energy}' uses only the two high-energy bands. 
\end{itemize}
While it is naively preferable for the `Neighboring' model to include only the neighboring two bands, these are both contaminated by Ar line structures that are substantially brighter near Sgr A* than the nearby continuum.  In addition, when we fit the 3.5 keV photons using this background model, inspection of the covariance matrix reveals near degeneracy between fitting the normalization of the two neighboring sidebands in both Perseus and the Galactic Center.  We therefore (minimally) add the 4.2-4.5 keV band in order to provide at least one  clean (``line-free'') continuum sample.  It is important to note that this continuum template does not resemble a dark matter profile (Fig.~\ref{fig:continuum_sample}) while improving the fit dramatically in each system.

\subsection{Emission Line Templates}
\label{subsec:line_templates}

\begin{figure}
\begin{center}
   \includegraphics[width=0.95\textwidth]{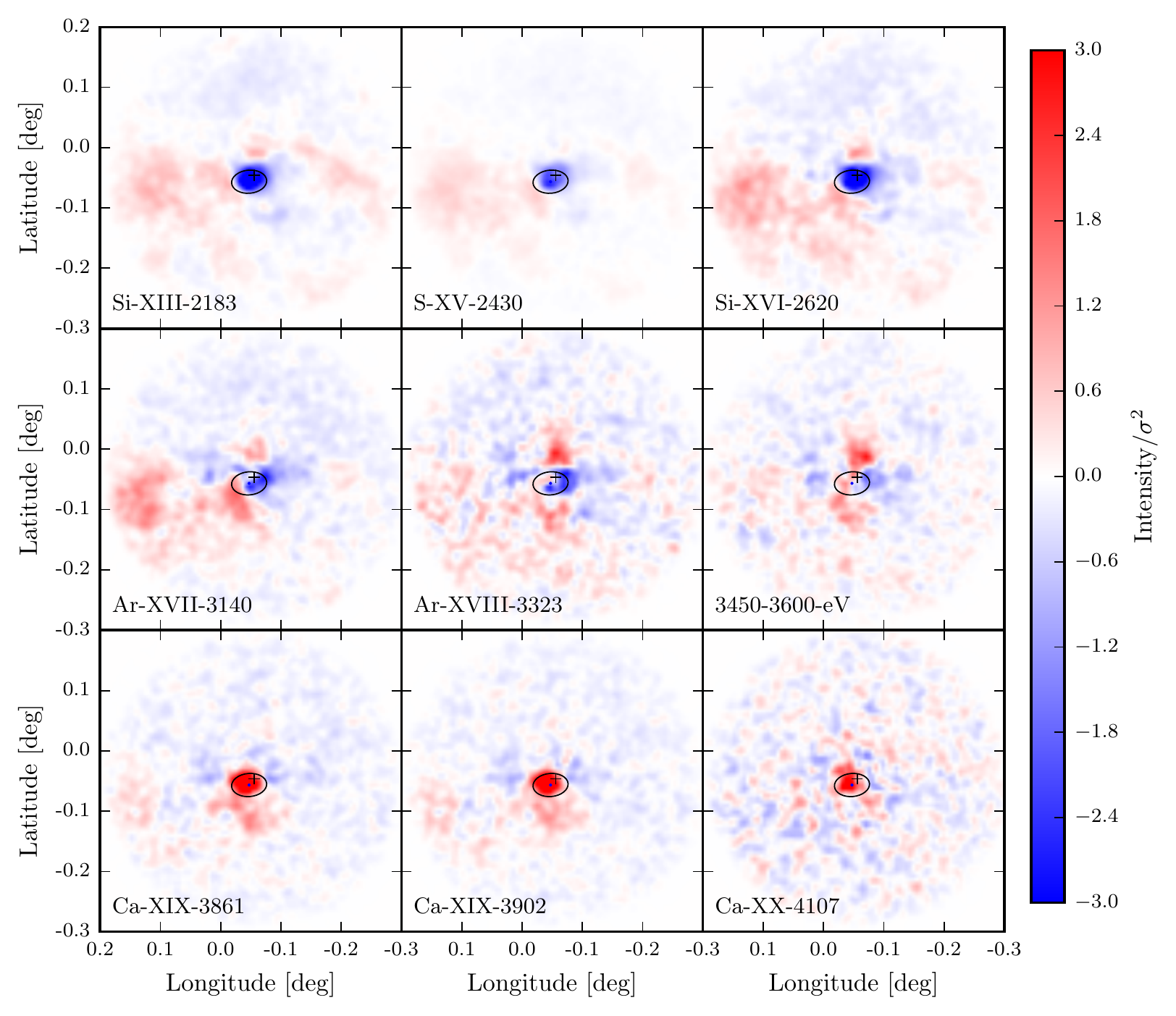}
   \caption{Morphology of 8 plasma emission lines including the 3.5 keV band surrounding the Galactic center region after subtracting off the best-fit (ML) contribution from 5 continuum bands.  For illustrative purposes, we normalize maps to the variance of each template.  The band from 3.45-3.6 keV is also shown in the center-right panel.  A black `+' indicates the location of Sgr A* while the outer shell of the supernova remnant Sgr A East is approximately bounded by the ellipse shown, from Ref.~\cite{Maeda:2002}.}
\label{fig:GC_lines}
\end{center}
\end{figure}

In order to search for qualitative correlations between the 3.5 keV exess and known plasma emission lines, we create templates for 8 spectral lines between 2 and 4.1 keV by selecting photons within $\pm50$~eV of each line's central energy\footnote{The Full-Width-Half-Maximum (FWHM) energy resolution of the MOS CCDs is approximately $100$~eV$\sqrt{E_\gamma/3.5~\rm{keV}}$. Performance characteristics for {\em XMM} used throughout this paper are taken from~\href{http://xmm.esac.esa.int/external/xmm\_user\_support/documentation/uhb\_2.1/node1.html} {http://xmm.esac.esa.int/external/xmm\_user\_support/documentation/uhb\_2.1/node1.html}.}. The complete list of lines we consider is in the first column of the left panel of Table~\ref{tab:templates}, along with the corresponding central energies and peak emissivity temperatures (the electron temperature for which the line intensity is maximal).  In Figure~\ref{fig:GC_lines} we show the line templates, as well as our 3.45-3.6 keV band of interest, with the best fitting `All' continuum model subtracted off in order to highlight the characteristic morphology of each line. A black `+' and ellipse indicate the location of Sgr A* and the shell of the supernova remnant (SNR) Sgr A East, from Ref.~\cite{Maeda:2002}. 

It is important to note that because some of the lines are weak with respect to the continuum emission, and because our continuum is slightly energy dependent, it is difficult to unambiguously separate `line photons' from `continuum photons'.  For most of the lines listed in Table~\ref{tab:templates}, the line flux is a substantial or even dominant fraction of the total emission, excepting the 3.45-3.6 keV band and some weaker lines in Perseus, which contribute only a few percent of the total flux. Later in our analysis we refer to `line' templates in two contexts: for qualitative comparisons we will subtract off the best fitting continuum model, and thus most of the residual should be characteristic of that spectral line.  When setting limits on the dark matter decay lifetime we include {\em all} photons in the respective line energy.  In each case where the line templates are used, one should keep the above caveats in mind, though we do not expect either the continuum or line emission to trace emission from dark matter.

Figure \ref{fig:GC_lines} shows a clear common feature observed across all lines whose emissivities peak at relatively low plasma temperatures, which exhibit a negative (oversubtracted) residual concentrated near SNR Sgr A East, which increases into a bright, $0.2^\circ$ diameter lobe extended toward positive galactic longitudes. In contrast, the lines which peak at higher plasma temperatures\footnote{We urge caution in strictly interpreting emission in terms of temperature, as varying relative elemental abundances also play an important role in the morphology.} are seen to be more centrally concentrated around Sgr A* and Sgr A East.  The 3.45-3.6 keV band shows a distinct quadrapolar morphology with positive residuals in the north/south oriented lobes.  Visually, this `residual' emission most closely resembles the Ar XVIII line at 3.32 keV. If this residual is representative of the line morphology (rather than mis-modeling of the energy dependent continuum), then it is evident that this emission profile is qualitatively incompatible with that expected from dark matter. We quantify this statement accurately in what follows. 

\begin{figure}[ht]
\begin{center}
   \includegraphics[width=0.95\textwidth]{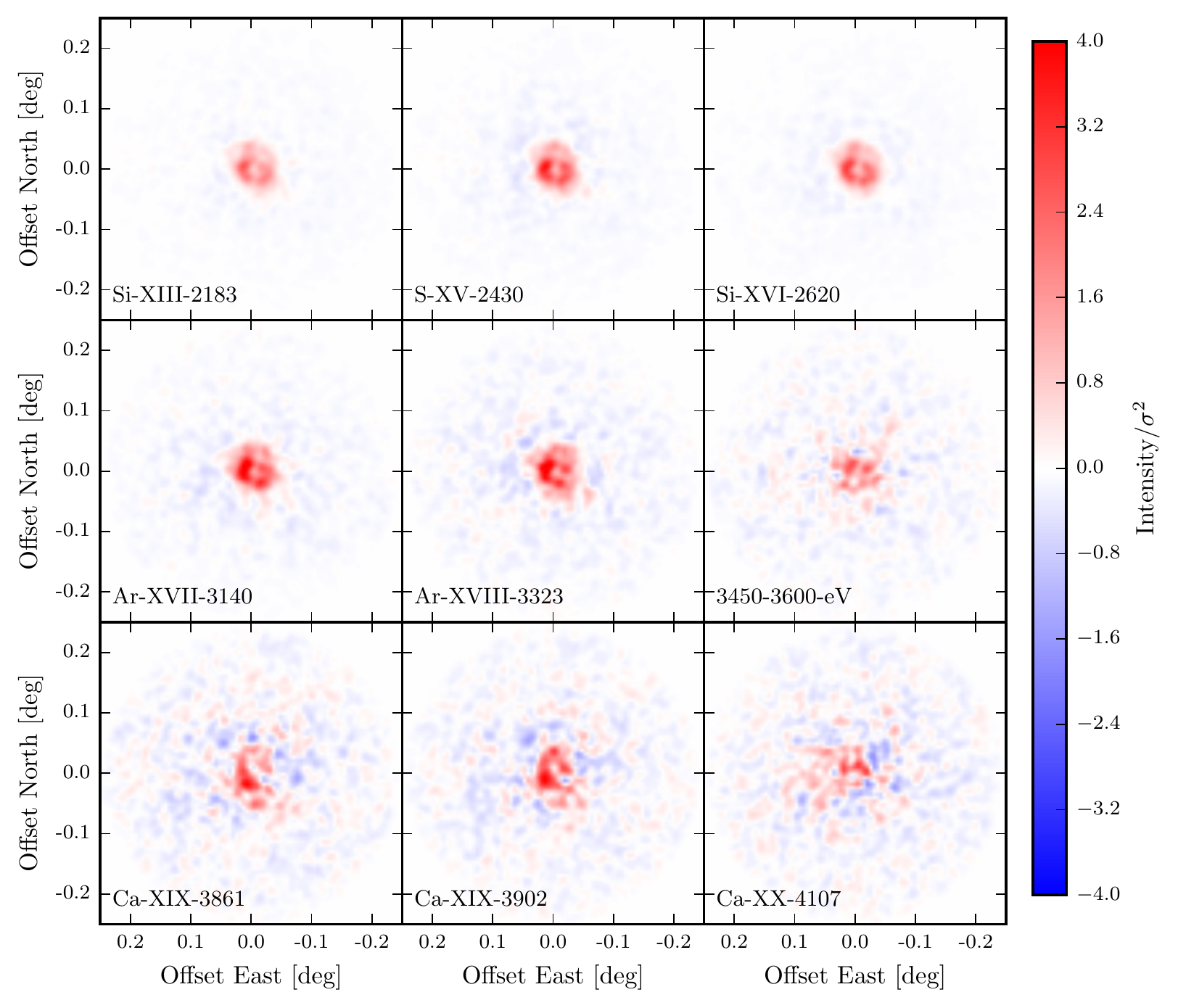}
   \caption{Morphologies for the 8 plasma emission lines and the 3.5 keV band in the Perseus cluster after subtracting off the best-fit (ML) contribution from 5 continuum bands.  The normalization prescription is as in Fig.~\ref{fig:GC_lines}.}
\label{fig:Perseus_lines}
\end{center}
\end{figure}

In Figure~\ref{fig:Perseus_lines}, we exhibit the same set of lines for the Perseus cluster.  At our binning resolution and exposure depth, many of the detailed asymmetric features in the cluster are unfortunately washed out, and one is left with an approximately spherical distribution of X-rays for both the continuum and the lines.  Still, the two well-known `ear-like' features (see e.g. Ref.~\cite{Fabian:2002}) are visible in the central regions, keeping in mind that point source masking and exposure maps induce some distortions.  Clearly, cooler emission lines are more concentrated in the cool cluster core, with a characteristic radius $\lesssim2.5'$, while emission lines associated with hotter, higher-dispersion plasmas extend well beyond the core.  The 3.5 keV band appears, here, to be associated both with the core, perhaps from residual low-energy continuum emission, as well as exhibiting a small extended component tracing the morphology of emission lines that peak at higher temperatures~($T_{\rm peak} \gtrsim 3~$ keV).  It is important to note that in the case of Perseus, the 3.5 keV line is (even globally) quite weak.  Thus a large, or even dominant portion of the residual emission in this band could be related to errors in continuum modeling.  In Section~\ref{subsec:residuals} we discuss potential systematic effects associated with the continuum model for the 3.5 keV band.

\subsection{Dark Matter Templates}
\label{subsec:dm_templates}

The differential flux of photons from dark matter decay factorizes into two components: one containing the particle physics (P. P.) factors and one associated with the particle-physics-model-independent astrophysics. 

\begin{align}
\frac{d\Phi_\gamma}{dE_\gamma}=&
\underbrace{ 
\frac{1}{4\pi}\sum_f \frac{\Gamma_f}{m_\chi} \frac{dN_\gamma}{dE_\gamma}
}_{\rm P.P.}
\underbrace{
\int_{\Delta\Omega}\int_{l.o.s} \mathcal{A}(l,b)\cdot\rho(l,b,z)~d\Omega dz
}_{\mathcal{J}{\rm -factor}}
\label{eq:decFlux}
\\[-2\baselineskip] \tag*{Decay}
\end{align}\\[-\baselineskip]

In the above equations, the first terms represent the particle physics for final-states $f$, dark matter mass $m_\chi$, and model-independent decay rate $\Gamma\equiv1/\tau$.  The modification of Eq.~(\ref{eq:decFlux}) above for the case of annihilation is trivial. In our primary model of interest, a sterile neutrino decays at loop level to a standard neutrino, radiating a nearly monochromatic photon with energy $E_\gamma\approx m_\chi/2$ in the process. The lifetime is independent of the particular sterile neutrino model and is given in terms of a mixing angle $\theta$ with active neutrinos~\cite{Pal:1981rm}.
\begin{align}
\tau = 7.2 \times 10^{29}~{\rm s}~ \left(\frac{10^{-4}}{\sin{(2\theta)}}\right)^2\left( \frac{1~{\rm keV}}{ m_\chi} \right)^5.
\label{eqn:neutrino_mixing}
\end{align}

The double integral term in Eqn.~\ref{eq:decFlux} is known as the $\mathcal{J}$-factor, with units GeV~$\rm{cm}^{-2}$ for decaying dark matter. The mask $\mathcal{A}(l,b)$ represents an efficiency factor due to gaps between CCDs, masked point source regions, defective pixels, and the off-axis effective area. The mask $\mathcal{A}$ takes values between 0 to 1 and is generated by summing the individual aligned observation and exposure masks, weighted by their Good exposure Time Intervals (GTIs).  Note that $\mathcal{A}$ itself is also an isotropic emission template which can be used to subtract the contribution of isotropic backgrounds such as galactic foreground and extragalactic background emission which are focused through the telescope.  For a cosmic-ray induced isotropic background template $\mathcal{A_{\rm flat}}$, we can use $\mathcal{A}$ without exposure masking (but including point-source and chip-gap masks), since these events are not focused through the telescope.  The majority of each of these isotropic events should be compensated by the sideband templates. However, this couples the relative isotropic normalizations to the continuum emission. We therefore include these templates when explicitly indicated, although this has little quantitative impact in practice. 

For the Galactic center field of view, we take three dark matter profiles including an NFW~\cite{NFW:1996}, an  Einasto profile~\cite{NFW:2004}, and the paradigmatically cored Burkert~\cite{Burkert:1995} profile, which serves as a conservative (low) estimate of dark matter emission from the inner Galaxy.  The functional form of the dark matter density profiles are defined as follows:
\begin{figure}
\begin{center}
   \includegraphics[width=0.95\textwidth]{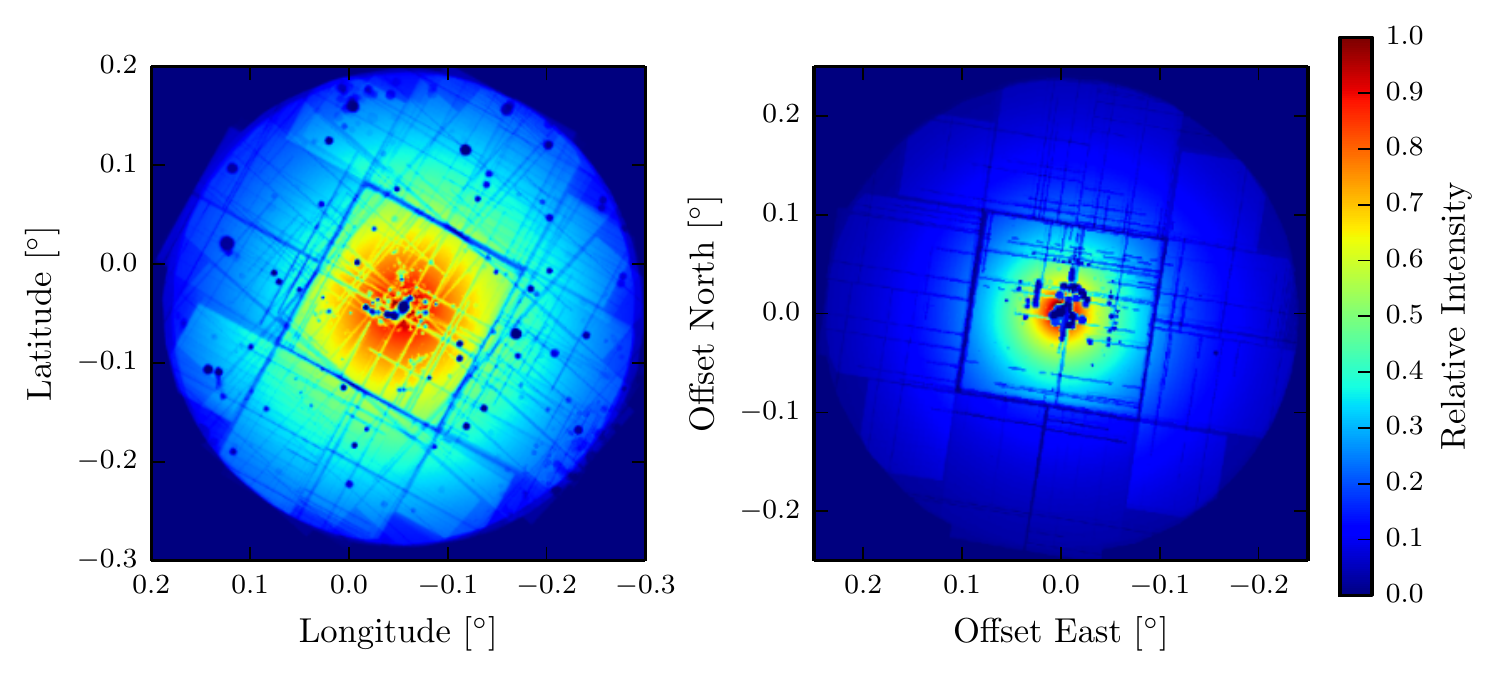}
   \caption{$\mathcal{J}$-factors sky-maps for a decaying dark matter candidate distributed as an NFW profile convolved with the \emph{XMM} MOS instrument response functions and exposure masks for the Galactic center field (left) and the Perseus cluster (right).}
\label{fig:example_skymaps}
\end{center}
\end{figure}
\begin{align}
\rho(r)&=\left(\frac{r_s}{r}\right)^{\alpha} \frac{\rho_0}{(1+r/r_s)^{3-\alpha}},\\[-.7\baselineskip] 
\tag*{NFW} \\
\rho(r)&=\rho_0 \exp{\left( \frac{-2}{\alpha}\left[\left(\frac{r}{r_s}\right)^{-\alpha}-1\right]    \right)},\\[-.7\baselineskip] 
\tag*{Einasto} \\
\rho(r)&= \frac{\rho_0~r_s^3}{(r_s+r)(r_s^2+r^2)},\\[-.7\baselineskip] 
\tag*{Burkert} 
\end{align}
where the scaling radius $r_s=\{20,20,6\}$~kpc, $\alpha=\{1,.16,{\rm N/A}\}$, and the normalization factors\footnote{The overall normalization is obviously irrelevant for our template analysis.} are chosen such that the local dark matter density $\rho(R_\odot)=0.4~$GeV~cm$^{-3}$ is reproduced at the solar radius $R_\odot=8.5$ kpc.

For the Perseus cluster we employ an NFW profile, with inner slope $\alpha=1$ and a scaling radius derived using the mass-concentration relation of Ref.~\cite{Buote:2007},

\begin{equation}
c_{\rm vir}=9 \left( \frac{M_{\rm vir}}{10^{14} M_\odot h^{-1}}\right)^{-0.172},
\end{equation}

where $c_{\rm vir}=R_{\rm vir}/r_s$. The virial overdensity is taken to be 200, with $M_{200}=10.8^{+.46}_{-.41} M_\odot h^{-1}$ and $r_{200}=2.66\pm 0.04$~Mpc~\cite{Reiprich:2002}, yielding NFW parameters $r_s=445\mp3~{\rm kpc}$ and $\rho_0=0.0217\pm.001$~GeV~cm$^{-3}$.  Following Ref.~\cite{Boyarsky:2014jta}, we assume that Perseus lies at redshift $z=0.0179$ ($d=72$~Mpc).

Once the $\mathcal{J}$-factors are calculated for each pixel, the skymap is smeared by the \emph{XMM} MOS point-spread function (PSF), taken to be a Gaussian with $\sigma=15''$, and normalized according to the total GTI of the field and assuming an on-axis effective area of 750~cm$^2$ for MOS1+MOS2 which is corrected off-axis using generated exposure maps. In Figure~\ref{fig:example_skymaps}, we show example templates for decaying NFW dark matter in the Galactic center and in the Perseus cluster. 

In addition to the instrumental effects discussed above, one may worry about the effects of X-ray absorption by the dense interstellar medium near the Galactic center.  In order to assess the impact of this, we have utilized high-resolution velocity integrated\footnote{In principle, one should not include absorption behind the GC region. Unfortunately, converting velocity cubes into positional space is extremely difficult in the direction of the Galactic center as the velocity relative to the Sun vanishes along this line of sight.  Accurate subtraction of radio continuum emission is also problematic due to bright point sources in the GC region.  A detailed study of the HI distribution toward the Galactic center is beyond the scope and necessity of the present analysis.} radio observations from the ATCA HI Galactic center survey~\cite{ATCA_GC}.  One can then relate the integrated line brightness to the HI column density, ${\rm N_H}$ (see e.g. Ref~\cite{Dickey_Lockman:1990}).  We then employ the absorption model of Wilms et al~\cite{Wilms:2000} which provides an optical depth per atomic hydrogen column density assuming elemental abundances typical of the ISM.  At 3.5 keV, the ISM is optically thin for all but the highest column densities (${\rm N_H\gtrsim10^{23}}$) and the maximal absorption in our region of interest is found to be less than 10\%, averaging only a few percent.  In addition, the morphology is not disk-like and actually slightly enhances the quadrapolar structure observed in Fig.~\ref{fig:GC_lines}.  While molecular hydrogen column densities can be even larger in the Galactic center and more concentrated in the Galactic plane, the photoionization cross-section for ${\rm H_2}$ at 3.5 keV is completely negligible relative to heavier elements~\cite{Wilms:2000}.  As a final check, we recomputed the dark matter limits found in section~\ref{subsec:limits} using this absorption map, finding only percent level changes in all cases.  For these qualitative and quantitative reasons, we neglect absorption throughout the rest of this analysis.

We have empirically derived spatial templates for a variety of line and continuum emission in the Galactic center and Perseus fields of view.  We have also  generated emission profiles for dark matter templates which have been convolved with the necessary instrument response functions and analysis masks.  All templates used are summarized in Table~\ref{tab:templates}. In the case of emission line templates we also include the electron temperatures for which the corresponding transition line is at its peak intensity~\cite{apec}.  In the next section we use the pixel-by-pixel binned likelihood analysis in order to determine the primary emission template components associated with the 3.45-3.6 keV band.
\begin{table}[tb]

\begin{center}
\small
\begin{tabular}[t]{lcc}
\hline
\hline\\[-2.4ex]
 Template    & E (keV)              &   T$_{\rm pk}$($10^7$ K)\\
\hline 
\hline\\[-2.4ex]
 Si XIII &  2.183          &                   1.0 \\
 S XV & 2.430             &                   1.26 \\
 Si XVI & 2.620           &                   2.0 \\
 Ar XVII & 3.140          &                   2.0 \\
 Ar XVIII & 3.323         &                   3.98\\
 Ca XIX & 3.861           &                   2.51\\
 Ca XIX & 3.902           &                   3.16\\
 Ca XX & 4.107            &                   5.01\\
\end{tabular}
\hspace{2mm}
\begin{tabular}[t]{lcc}
\hline
\hline\\[-2.4ex]
 Template       &  E (keV)         &   \\
\hline
\hline\\[-2.4ex]
 Ctnm1 & 3.190-3.270        &               \\
 Ctnm2 & 3.373-3.450        &               \\
 Ctnm3 & 3.600-3.811$^{\dagger}$  &         \\
 Ctnm4 & 4.200-4.500        &               \\
 Ctnm5 & 4.500-4.800        &               \\
 DM  & --             &               \\
 Focused Isotropic  & --            &                \\ 
 Flat Isotropic  & --            &                \\ 
 \end{tabular}
\end{center}
\caption{Summary of templates used.  For atomic emission lines (top), we also indicate the peak emissivity temperatures taken from the {\tt AtomDB 2.0.2}~\cite{apec}. We note that the K XVIII emission peaks at 2.16 keV ($2.51\times10^7$~K). $^\dagger$This continuum band includes three weak lines from Ar XVII;  the limited energy resolution of the EPIC MOS CCD's prevents the lines from being cleanly separated and they are thus treated as a single broad continuum band.}
\label{tab:templates}
\end{table}

\section{Results} 
\label{sec:gc}

In this Section, we discuss the basic morphology of the 3.45-3.6 keV residual emission using each of our three continuum model choices (sec~\ref{subsec:residuals}).  This involves subtracting off the best-fitting continuum+isotropic templates and comparing the results against expectations for dark matter and line-like emission.  Next, in sec.~\ref{subsec:limits}, we add a decaying DM template to the fit, finding that there is no indication for a component whose morphology follows the dark matter templates in either the Galactic center or the Perseus cluster.  We then calculate upper limits on the particle decay lifetime and discuss the implications for a dark matter interpretation of the 3.5 keV line. Finally, we use the maximum likelihood technique to check for spatial correlations between the 3.5 keV band and nearby energies, uncovering strong statistical evidence in favor of an astrophysical interpretation of the 3.5 keV line (sec.~\ref{subsec:window}).

\subsection{Residual Maps}
\label{subsec:residuals}

\begin{figure}
\begin{center}
    \subfigure{\includegraphics[width=\textwidth]{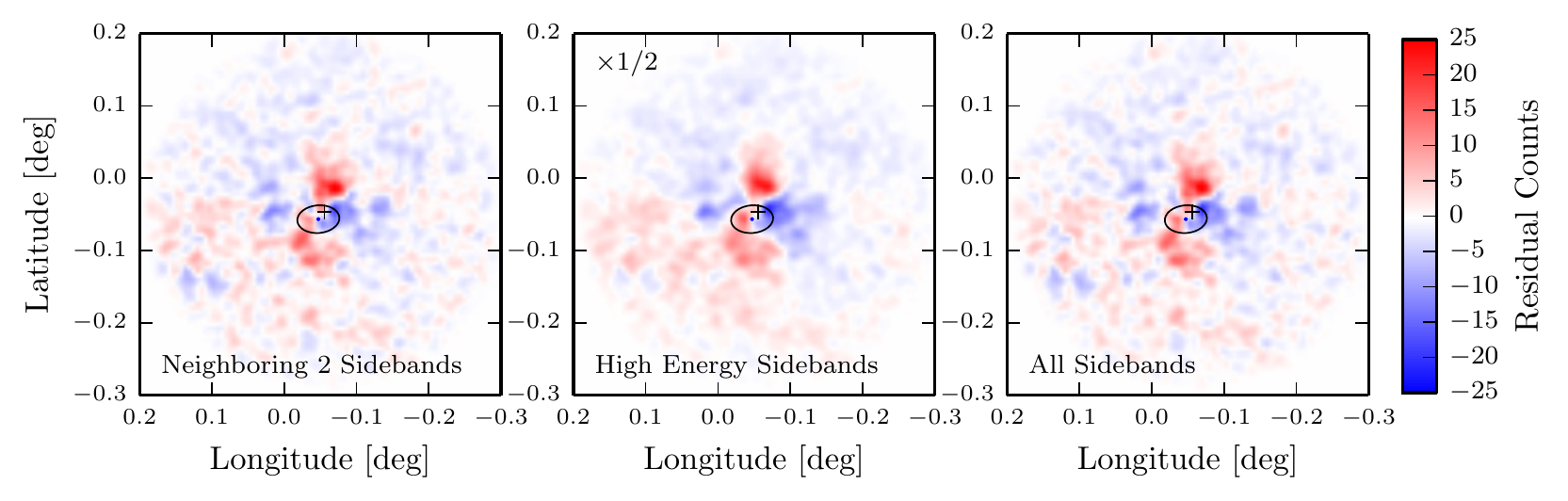}}\\\vspace{-4mm}
    \subfigure{\includegraphics[width=\textwidth]{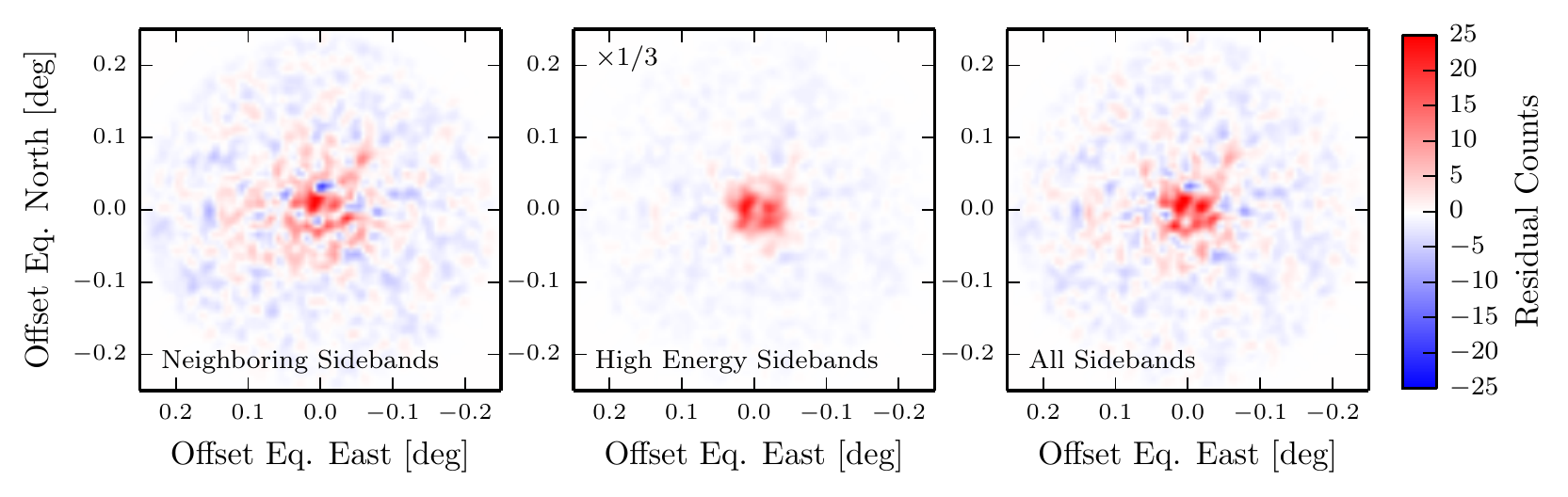}}
\end{center} 
    \caption{Comparisons between 3.45-3.6 keV residuals after subtracting the best fit continuum+isotropic templates for different models of the continuum near the Galactc Center (top row) and the Perseus cluster (bottom row).  Black `+'s in the top row indicate Sgr A* while the shell of SNR Sgr A East is shown by the black ellipse~\cite{Maeda:2002}.  Maps have been smoothed by a Gaussian kernel with $\sigma=20''$.  For the high-energy continuum model, the GC and Perseus cluster maps have been rescaled by a factor 1/2 and 1/3, respectively, in order to maintain visibility on a common scale.}
    \label{fig:residuals}
 \end{figure}

We perform a maximum likelihood analysis of the Galactic center and of the Perseus cluster X-ray observations listed above for energies between 3.45-3.6 keV, subtracting off the best-fitting linear combination of (1) isotropic templates, and (2) each of the `All', `Neighboring', and 'High-Energy' continuum template sets described in Section \ref{sec:continuum}.  Figure~\ref{fig:residuals}  shows the residual count maps for the Galactic center (top panels) and for the Perseus cluster (bottom panels) for each of the three different continuum models.

For each continuum model in the GC we observe a bright extension emanating roughly perpendicularly to the Galactic plane from the position of Sgr A* (black `+'), with a smaller bridge connecting toward the direction of the supernova remnant Sgr A East (the outer shell is represented by the black ellipse~\cite{Maeda:2002}) as well as a large faint bubble extending to the positive longitude edge of the field.  Although this residual appears most likely associated with Sgr A* or Sgr A East, the central few parsecs are not well resolved and are known to host a large number of potential sources. The largest residual is seen when using the high-energy model, indicating that this feature is at least partially aligned with the $\approx 3.2$~keV and $\approx 3.5$~ keV continuum.  In the `All' and `Neighboring' models, the residual is substantially reduced and the log-likelihood is greatly increased.

In the Perseus cluster (Fig.~\ref{fig:residuals}, bottom), prominent morphological identifiers are seen in each residual.  A relatively bright core region is apparent. Extensions to large radii are also observed, similar to those observed in the hotter Calcium lines, and to a lesser extent in Ar XVIII.  The clumped nature of this residual is difficult to reconcile with the much smoother distribution expected from dark matter as is the radial profile which has a much sharper gradient at the edge of the core than what expected from a decaying dark matter profile (cf. Fig.~\ref{fig:example_skymaps}).

\begin{figure}[width=\textwidth]
	\begin{center}
    \subfigure{\includegraphics[width=.49\textwidth] {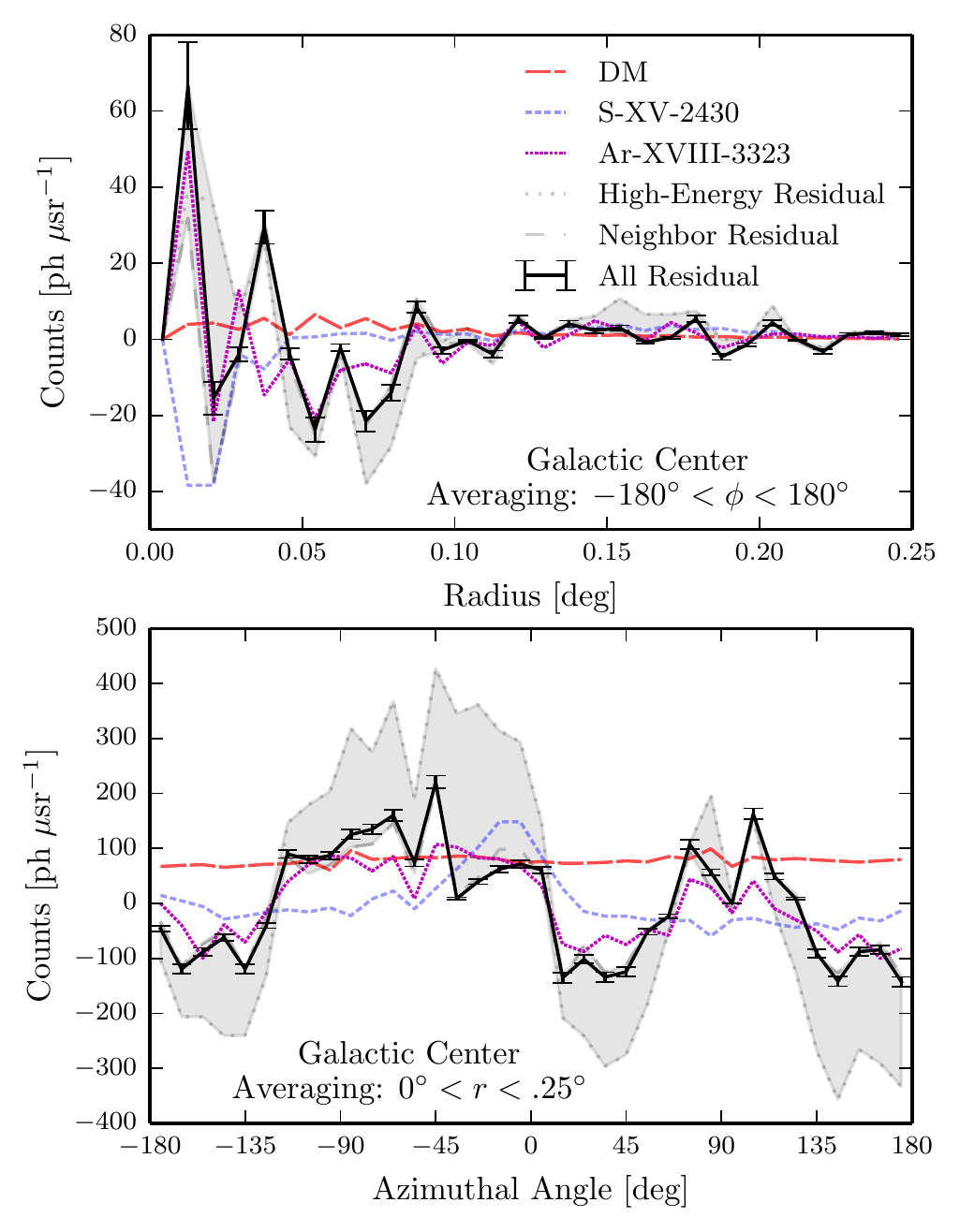}}
    \hspace{-2mm}
    \subfigure{\includegraphics[width=.49\textwidth] {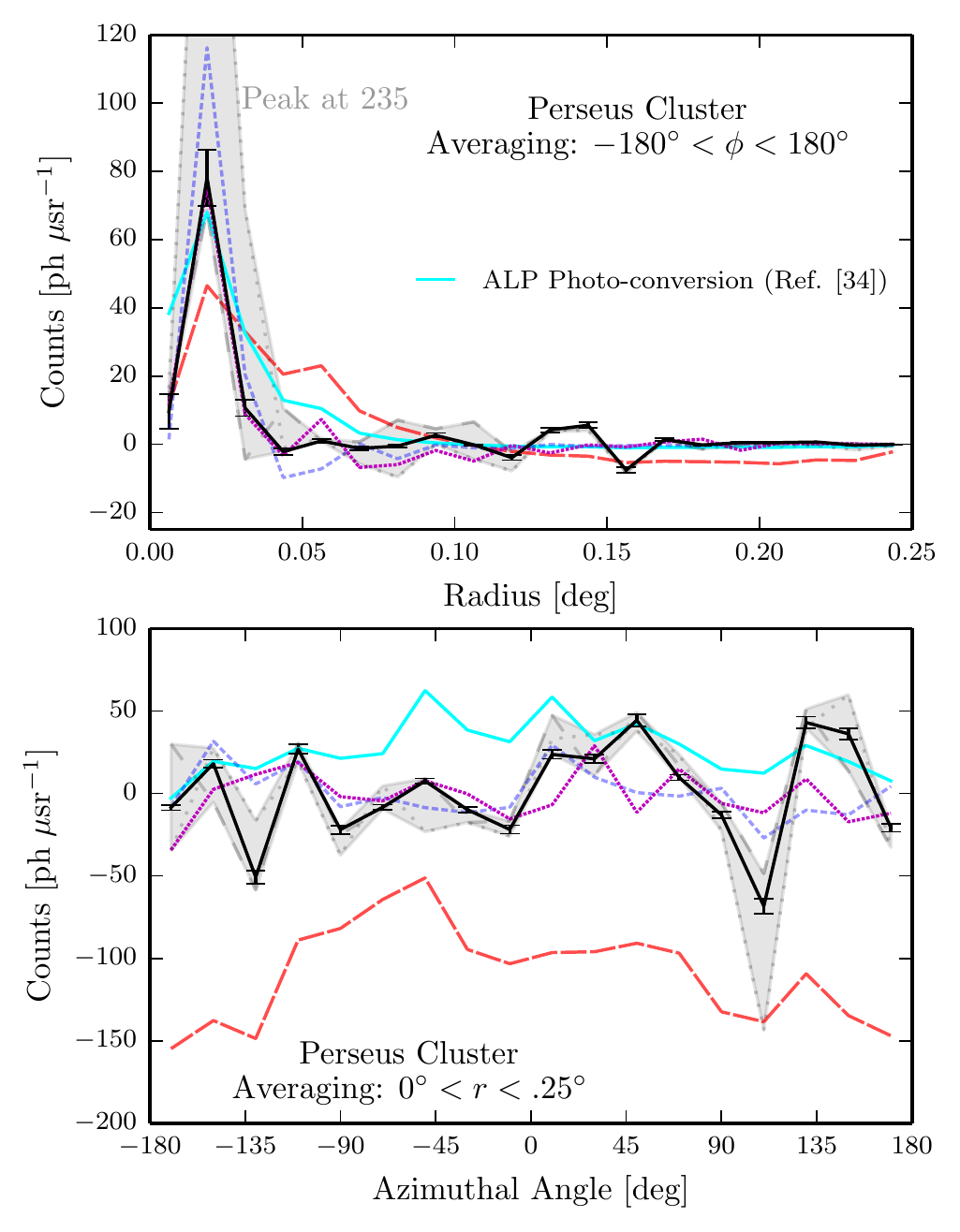}}    
    \end{center}
   \caption{Radial and azimuthal profiles for the Galactic center (left panels) and for the Perseus cluster's (right panels) un-smoothed residual maps shown in Fig~\ref{fig:residuals}.  The shaded regions bracket alternative continuum models along with the best fitting NFW template (red), S XV Line (light blue), and Ar XVIII line (magenta).  Poisson error bars are shown for our `All' model.  Azimuthal profiles rotate clockwise from the line pointing from the center (Sgr A* in the GC) to positive longitudes.  In the top-right panel, we also show the steepest radial profile expected from photo-conversion of axion-like-particles (solid blue), calculated using formulae in Ref.~\cite{Conlon:2014wna} and convolved with the relevant masks and instrument response.}
  \label{fig:residual_profiles}
\end{figure}

In Figure~\ref{fig:residual_profiles} we show averaged radial and azimuthal intensity profiles for each of the residuals along with the NFW DM templates, a prototypical cool emission line (S XV at 2.43 keV), and a medium-temperature emission line (Ar XVIII at 3.32 keV), both with the `All' continuum model already subtracted.  The normalizations of lines and DM templates are determined by minimizing the total $\chi^2$ of the binned template in question to the `All' residual -- i.e. the best fit is to the entire template rather than fitting the averaged profiles individually.  Because the dark matter templates are positive everywhere, we have also included a flat isotropic degree of freedom in this case.  

By visual inspection, it is evident that standard decaying dark matter does not provide a good fit to either the GC or Perseus cluster system azimuthally or radially, while the Ar XVIII line morphology shares precisely the same features as the GC residual and both the Ar XVIII and the S XV profiles share common features with the Perseus residual.  The GC is characterized by a distinctly quadrupolar profile in azimuth and by a centrally peaked radial profile.  The DM template is instead effectively flat in azimuth, with a slow radial falloff. Notice that the small deviations seen in the DM template are due to the combined exposure masks $\mathcal{A}(l,b)$.  As noted in Sec.~\ref{subsec:dm_templates}, absorption effects are negligible and are, even qualitatively, unable to account for such a morphology. The Perseus residual shows evidence of a distinct core which is truncated around $0.04^\circ\approx 50~$kpc, matching what seen for the lines.  Compared with DM, this is much too sharply peaked than even the most concentrated physical NFW profile.  

While the Ar XVIII line shown is detected at relatively high significance in the GC and Perseus cluster, it is substantially weaker than that of Ar XVII. The Ar XVII line morphology is similar to Ar XVIII except at the far East side of the field where a large excess is present in the GC. One must consider that a portion of any `line' template will likely be unmodeled continuum emission.  This is especially true in case of weak lines, where the continuum dominates the total flux. However, the foremost objective of this analysis is to test whether the excess emission in the 3.5 keV band traces {\em any} plausible astrophysical background, and whether dark matter can potentially provide a satisfactory spatial morphology.  These questions will be answered quantitatively in the following sections.  

For Perseus, we also overlay the steepest of the radial profiles expected from photoconversion of axion-like-particles (ALPs) in Perseus' large scale magnetic field as calculated using Eqns. (3.1) and (3.3) of Ref.~\cite{Conlon:2014wna}, with a free electron density $n_e$ taken from Ref.~\cite{perseus_free_electron}, and using the NFW parameters specified in our subsec~\ref{subsec:dm_templates}. The three dimensional profile for the axion signal is then proportional to $\rho(r)\times n_e(r)^{2\eta}$ and is always steeper than the decaying DM case for a radially decaying magnetic field, as is measured in Perseus.  The case shown corresponds to the steepest magnetic field profile, $\eta=1$, while smaller values of $\eta$ lead to a significant flattening. The projected skymap was then convolved with the relevant masks.  

The ALP scenario is significantly steeper than the decaying DM case due to the magnetic field falloff and it visually appears marginally compatible with the morphology of the residual emission. As $\eta\to0$, this profile asymptotes to the decaying DM case.  The azimuthal profile is also reasonably compatible, though much less so than for the  profiles corresponding to elemental lines.  However, here we have used an idealized model for the magnetic field structure which in reality will be much more complicated and could follow an azimuthal profile similar to that of emission lines. While the steepest ALP-conversion profile morphologically traces the cluster's core, the excess emission is much better fit by adding a continuum or low energy line template (after which there is no preference for ALPs). We discuss this aspect on more quantitative grounds in the next section.

Recently, Ref.~\cite{Conlon2014:GC} also calculated the morphology expected in the Milky Way's center where the signal is expected to roughly trace the projected free electron density (see also Ref.~\cite{Conlon:2014xsa}).  Based on the NE2001 model for the free electron distribution~\cite{NE2001:I,NE2001:II}, the expected signal is (i) highly elliptical with an axis ratio of nearly 4:1 elongated in the Galactic plane, and (ii) has a peak intensity offset from the center $\approx20'$ toward Galactic north.  Neither of these features are remotely compatible with the observed excess.  In fact, the orientation of an ALP-conversion excess is orthogonal to the residual shown here making it maximally incompatible with observations.  A successful interpretation of the 3.5 keV line in terms of ALP photoconversion seems unlikely across both objects simultaneously.  The case for an astrophysical origin of the GC 3.5 keV signal thus appears very robust.

In both systems, the 3.5 keV line makes up only a few percent of the total flux.  An important caveat to above discussion is therefore to understand whether these residuals are representative of the actual 3.5 keV signal, as opposed to an energy dependent feature of the continuum that is not included in our model.  The above is intended as a qualitative comparison, but in the following two sections we will draw the same conclusions, observing a stronger statistical correlation between the 3.5 keV band and, for example, the Ar-XVIII line shown in Figure~\ref{fig:residual_profiles}, than continuum regions. Note that in what follows we do not derive limits or perform template fitting to the residuals above, but instead perform new fits allowing the normalizations of all of the included templates to vary, minimizing the impact of errors in the continuum model.

\subsection{Fits and Limits for Decaying Dark Matter}\label{subsec:limits}

In this section we test statistically for the presence of a dark matter component in the {\em spatial} distribution of photons in the 3.45-3.6 keV band.  Using the method of maximum likelihood  described above, our model consists of up to eight templates: one for dark matter, one to two for flat and focused isotropic templates, and up to five more, depending on the continuum model.  These template normalizations are then varied to fit to the all photons in the 3.45-3.6 keV band -- i.e.~{\em not} to the residuals found in the previous section -- and similarly when calculating upper limits. Our key finding is that {\em there is no statistical evidence (TS$\approx$0) for decaying dark matter in the Galactic center region for any combination of the three DM halo models or continuum models, with or without any combination of isotropic templates}.

As additional tests of the robustness of our results, we start with the `All' continuum and an `NFW' profile and allow the central location of the DM profile to float in the vicinity of Sgr A*, increase the size of the spatial binning, radially mask the inner 5 arcminutes, and mask the outer 5 arcminutes, finding that each variation produces no change in the (zero) statistical significance of the dark matter template.  Finally, we scan the inner slope of the NFW profile between 0 and 3, finding no preferred value.  Since the DM profile is a pure power-law this close to the GC, annihilating dark matter -- with the intensity canonically tracing $r^{-2}$ -- is also essentially ruled out as a candidate explanation of the 3.5 keV excess from the GC at high confidence level.  We reiterate that models where the signal is azimuthally flat, including e.g. eXciting Dark Matter \cite{Finkbeiner:2014sja} or annihilating dark matter scenarios (e.g. \cite{Dudas:2014ixa}), are clearly disfavored by the quadrupolar structure of the residuals from the GC.

Similarly, the morphology of the X-ray emission from Perseus also does not exhibit any evidence for a dark matter component for the default cluster NFW model using any continuum model.  Allowing the inner slope and scale factor to vary does not increase the goodness of fit, unless the relevant parameters are set to completely unphysical values (e.g. $r_s
=60$~kpc and $\gamma=1.6$ compared to 445~kpc and 1 respectively), in which case there is a slight statistical preference.  These parameters effectively try to mimic the profile of the cool inner core, making it clear that the excess emission is associated with unmodeled cool lines or with thermal continuum emission below 3 keV.  We will explore this further in sec.~\ref{subsec:window}.  

We also test the case of ALP photoconversion for Perseus.  Using the `All' continuum model and both isotropic templates we find a TS=28 preference for the steepest ALP profile (corresponding to $\eta=1$ in subsection~\ref{subsec:residuals}).  We note that if the 3.5 keV line is due to K XVIII, we expect to see strong correlations with the bright thermal emission below 3 keV.  This continuum is not included in our three continuum models, and if we add continuum and/or line templates below 3 keV, these templates pick up TS$>$100 and the ALP template's TS becomes insignificant.  The significance is also reduced for the shallower $\eta=0.5$ scenario.  While we cannot rule out an ALP component with robust statistical significance due to the inherent morphological similarity to the core, it is clear that an elemental origin is strongly preferred.  This is also a natural interpretation given the strong incompatibility of the ALP photoconversion scenario with an excess in the Galactic Center. 

We now calculate upper limits on the dark matter decay rate and sterile neutrino mixing angle.  While what was described above has focused on spatially correlating the 3.5 keV excess with nearby regions of the X-ray spectrum, such analyses carry the important caveat that they may be sensitive to the choice of background models.  If the templates employed in the fit -- or some combination of them -- are able to mimic the emission profile of dark matter, then there would be no statistical preference for adding a (now redundant) DM template. When placing upper limits, the story is more subtle. On the one hand, as the normalization of the dark matter template is increased, additional degrees of freedom work to accommodate the `imposed' DM component. On the other hand, if the fit is poor to begin with, a small number of critical pixels may be degenerate with a DM template; for example, within the brightest inner arcminutes of Sgr A*.  In light of these competing effects we explicitly check each template combination in order to systematically assess limits.  

In Table~\ref{tab:limits} we show the 95\% confidence level lower limits on the dark matter decay lifetime for NFW, Einasto, and Burkert profiles in the Galactic center, and for the expected NFW profile for the Perseus cluster using the maximal template set: the `All' continuum, both isotropic templates, and all lines.  These have been verified to be the most conservative\footnote{The sole exception is for the `Neighboring' continuum case, where the limits weaken by a factor $\sim 2$ in the GC.  However, if we mask the inner 2' of the DM mask, the limits become more stringent than for the `All+Lines' case, opposite of what would happen if a true dark matter component were present.} and have also been translated into upper limits on the sterile neutrino mixing angle using Eqn.~(\ref{eqn:neutrino_mixing}). Remarkably, our morphological analysis of the Galactic center provides the most stringent available limits.  Even in the most conservative GC case, corresponding to the Burkert profile, the upper limit on the mixing angle essentially rules out preferred values for M31 in Ref.~\cite{Boyarsky:2014jta}, while the other profiles rule these out at high significance.  The GC results also rule out a standard decaying dark matter interpretation of the stacked cluster analysis in Ref.~\cite{Bulbul:2014sua}, unless a radically different morphology is expected, as may be the case for e.g. photo-conversion of axion-like particles where emission depends also on the local transverse magnetic field structure.  As shown in the previous section, even this scenario is morphologically inconsistent with observations.

The flux expected from the Perseus cluster is significantly weaker and the limits are much less stringent.  In particular, the `All' case is fully compatible with measurements of M31 from Boyarsky et al~\cite{Boyarsky:2014jta}, but still rules out the detected mixing angle for the Perseus (MOS including core) analysis of Ref.~\cite{Bulbul:2014sua}.  The `All+Lines' case is the most conservative estimate and does not limit the parameter space of interest.  However, in contrast to the Galactic center case, the line templates for Perseus are broadly azimuthally symmetric, allowing them to more easily allow for a DM profile with an incorrect radial profile.  For this reason, we believe these limits are likely to be overly conservative.

\begin{table}
\begin{center}
\small
\begin{tabular}{lccccc}
\hline\\[-2ex]
Target   & Template Set   & Profile    &   $\mathcal{J}$ & $\tau$ & $\sin^2(2 \theta)$\\
&  &  & $(10^{18}~\rm{GeV~cm^{-2}})$ & $(10^{28}~s)$ & \\ 
\hline \hline\\[-2ex]
GC & All+Lines & NFW & 6.8 & $>$3.7 & $<1.1\times10^{-11}$ \\
GC & All+Lines & Ein & 8.2 & $>$5.9 & $<7.0\times10^{-12}$ \\
GC & All+Lines & Bur & 1.9 & $>$1.3 & $<3.3\times10^{-11}$ \\
Perseus & All & NFW & 1.4 & $>$.55 & $<7.5\times10^{-10}$ \\
Perseus & All+Lines & NFW & 1.4 & $>$.03 & $<1.5\times10^{-9}$ \\
\hline

	\end{tabular}
\end{center}
\caption{$\mathcal{J}$-factors for decaying dark matter profiles and limits on a sterile neutrino's lifetime and mixing angle the Galactic center and Perseus cluster \emph{spatial} analyses using the delta-log-likelihood method at 95\% confidence level.  Each of the line and continuum templates were used in deriving limits. These results exclude even the smallest value of $\sin^2(2\theta)$ from Ref.~\cite{Boyarsky:2014jta} at 3.4 and 4.7$\sigma$ for NFW and Einasto, respectively.}
\label{tab:limits}
\end{table}

\subsection{Cross-Correlating the 3.45-3.6 keV Emission}
\label{subsec:window}

To assess the level of spatial correlation between the 3.45-3.6 keV band and different spectral regions, we compare the likelihood of a continuum only fit with that of the continuum plus a narrow 50 eV-wide ``sliding window'' template, which is scanned over energy.  The test statistic of this template should then peak if the 3.5 keV band is correlated with all or some line emission in other regions of the spectrum. The null model used consists of the five `All' continuum bands along with two additional low-energy templates covering the ranges 2.23-2.38 and 2.68-2.78 keV\footnote{Due to the substantial overlap of these regions with some prominent low-energy spectral lines, we have not used them previously.  In what follows, however, their inclusion is a more conservative approach since any line emission `leaking' into the continuum model will result in a lower TS for templates centered on emission lines.}. Next we add one more template created using photons in a 50 eV wide window (slightly lower than the FWHM energy resolution of {\em XMM's} MOS sensors) and scan the central energy between 1.5-5 keV, finding the maximum likelihood at each point. 

\begin{figure}
	\begin{center}
    \subfigure{\includegraphics[width=.95\textwidth]{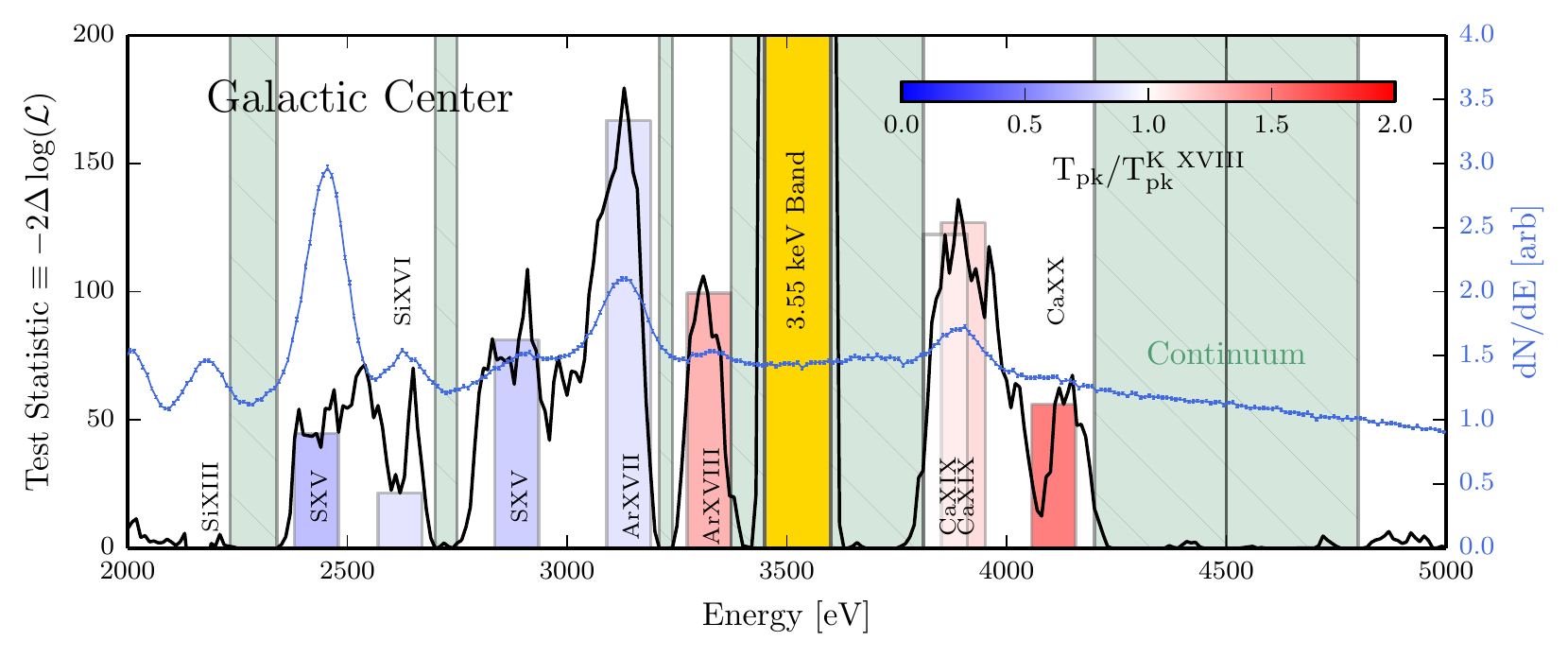}}
    \subfigure{\includegraphics[width=.95\textwidth]{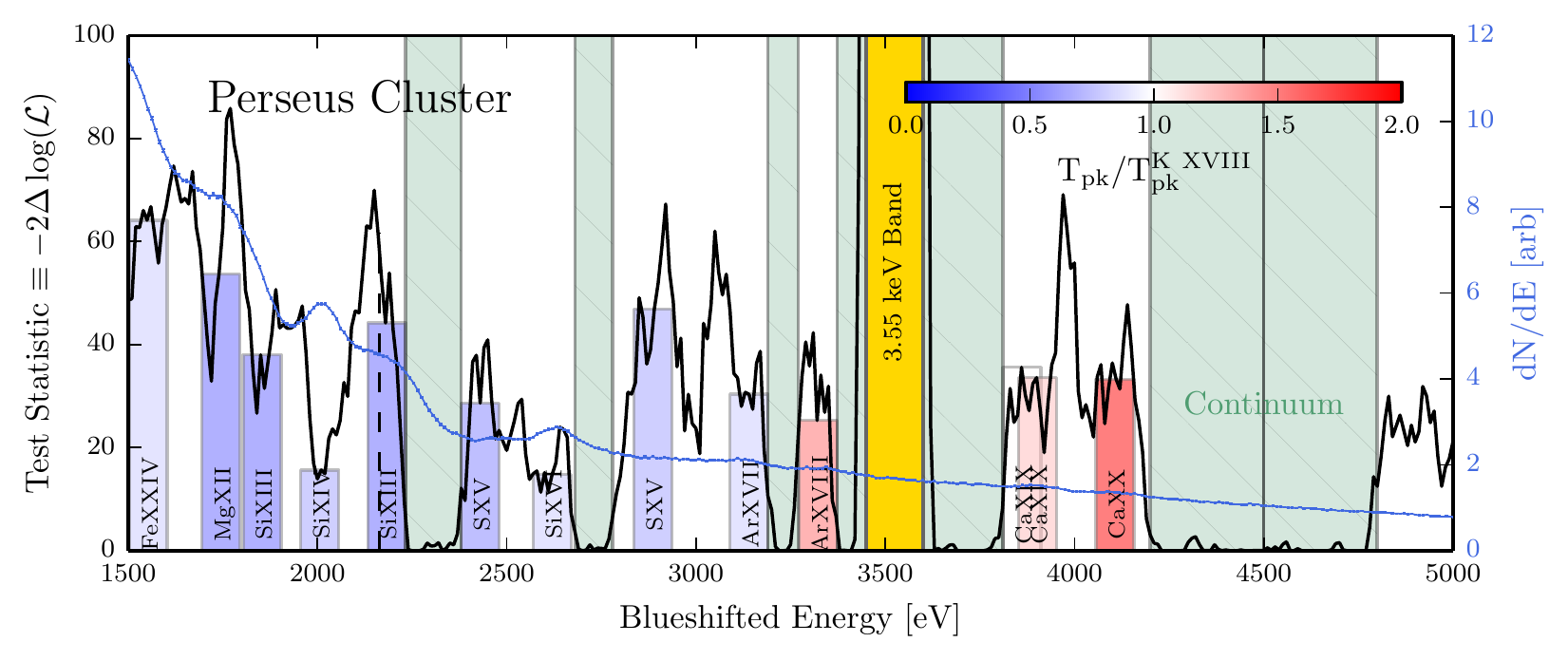}}
    \caption{Shown in black is the test statistics corresponding to adding a sliding 50 eV-wide window template to a null model consisting of 7 continuum bands (green hatched regions).  In light blue we also overlay the raw {\em XMM} spectrum for the Galactic center (top) and Perseus cluster (bottom).  The 3.45-3.6 keV band is highlighted in gold. The brightest spectral line templates (with correct widths) are color coded according to the ratio of a given line's peak emissivity temperature to that of K XVIII.  T$_{\rm peak}$ for K XVIII is also represented by a dashed line at 2.16 keV, visible in the Perseus plot.  For both the Gaclactic center and Perseus the TS stemming from adding a dark matter template is zero.}    
    \label{fig:decomp}
    \end{center}
 \end{figure}

In Figure~\ref{fig:decomp} we show the test statistic as a function of the window energy (black line) for the Galactic center and Perseus cluster along with an overlay of the raw spectra for each system (light blue line with errorbars). We also indicate the 3.45-3.6 keV band (gold), continuum bands (hatched green), and spectral lines from Section~\ref{subsec:line_templates} (and a few extra), color coded by the ratio of their peak emissivity (electron) temperatures to that of K XVIII taken from {\tt AtomDB 2.0.2}~\cite{apec}.  The width of each band corresponds to the included photon energy range while the height indicates the TS of the sliding window at the central energy of each line. The height of the continuum and 3.45-3.6 keV band are arbitrary.

Immediately, one can see a dramatic improvement to the Galactic center fit when the sliding template overlaps with any spectral line above 2.5 keV, peaking for the Ar XVII line at 3.14 keV, but also very high for the Ar XVIII and Ca XIX lines.  If the Poisson uncertainty were the only source of error, the corresponding significance would be more than 10$\sigma$ compared to TS=0 when adding a dark matter template.  Of course, the TS becomes very large when the window overlaps with the 3.45-3.6 keV band since the fit becomes nearly perfectly autocorrelated.  Contrarily, the TS runs to zero when the energy window overlaps with continuum bands already included in the fit, since little to no additional template ``information'' is added.  

The peaks of the line emission must be compared against neighboring pseudo-continuum regions which are not included in the null model.  For example, the fit improvement when the window is near 3 keV is much smaller than that of the neighboring spectral lines, indicating two things:  First, the continuum emission in the 3.5 keV band is already well modeled.  Since the continuum morphology changes rather slowly in energy, adding more continuum templates offers little improvement;  Second, for the Galactic center the morphology of the residual emission is strongly correlated with that of atomic transition lines.  Since we are using the nearby continuum as a reference point, this can only be attributed to the presence of a distinct emission component over a narrow energy range.  The cool lines below 2.5 keV are seen to have the opposite behavior, where the TS is instead reduced compared to the nearby continuum model.  This is indicative of a negative correlation with these spectral lines, and their template normalization is preferentially driven to zero.

For each emission line, there exists a unique electron temperature where the emissivity peaks.  If each element is assumed to be identically distributed, one would then expect a parallel morphology between emission lines that peak at similar temperatures.  Roughly speaking, this is observed here, where the very-cool and very-hot lines do not provide a substantially improved fit. It is difficult to quantify this further for several reasons.  Firstly, over the few keV energy range of interest here, the emissivity as a function of temperature changes by less than a factor two for the emission lines considered. Thus we have only weak sensitivity to the underlying plasma temperature using morphology.  Secondly, it is important to note that chemical abundances are not uniform over the physical scales probed here.  As noted by e.g. Ref.~\cite{Maeda:2002} in studying Sgr A East, heavy elements such as iron are distributed much more compactly compared with He-like elements.  More generally, recent simulations of supernova explosions show that hydrodynamic instabilities can lead to highly asymmetric ejection of heavy elements~\cite{Wongwathanarat:2013}.  These factors complicate, for example, showing a nearly perfect correlation between suspected K XVIII at 3.5 keV and a spectral line with identical peak emissivity temperatures while also showing anti-correlation with cooler and hotter transition lines.

In the lower-panel of Figure~\ref{fig:decomp} we show results for the Perseus cluster.  Here the situation is very different.  The relevant physical scale is increased by more than $10^3$ and emission is dominated by the approximately isothermal cool-core surrounded by increasingly warmer and fainter radial shells.  This is in contrast to the Galactic center, where the continuum flux is dominated by non-thermal emission (predominantly unresolved point sources).  On this scale, entire galaxies are nearly point-like and the relative distribution of heavy elements is averaged out, depending much less on the atomic species in question.  Emission lines above 2.5~keV are much weaker in Perseus, appearing only as small bumps in the spectrum. Still, distinct peaks are observed for the cooler Si XII and S XV lines while the Ar lines are more ambiguous.  The hot Ca lines are now anti-correlated compared to the in-between continuum region at 4 keV. 

Below 2.4 keV the flux rises very rapidly and it becomes difficult to disentangle the continuum and line emission.  We note that the K XVIII line at 3.515 keV peaks at a plasma temperature of $\approx$2.16 keV, indicated by a vertical dashed line. In the optically thin limit, the power spectrum of the corresponding thermal bremsstrahlung is exponentially suppressed above 2.16 keV and is relatively flat below, having at most a $T_e^{-1/2}$ dependence on temperature.  Since we would like to test for correlation with this very low temperature continuum, this is not included in our baseline model. The suppressed TS observed at Si line energies could therefore be interpreted as anti-correlated Si lines plus a strong continuum component due to a low temperature plasma.  Such a plasma must exist, as the core temperature of Perseus has been measured to have a strong component near $kT\approx2$~keV~\cite{Fabian:2002}.  We note that the elemental abundances of bremsstrahlung-generating ionized hydrogen and K XVIII are likely to be different.  The line structure is too dense below 2 keV to gain additional insight with such a simplistic analysis.

For the Perseus cluster, our key finding is thus that the strongest correlations are observed with the cluster's core, namely continuum emission below $\approx2.16$ keV and select lines that peak at similarly low electron temperatures.  Such correlation is expected if the emission is due primarily to an atomic transition line of K XVIII, and is {\em not} expected if the emission is due to dark matter.   The recent findings of Ref.~\cite{Urban:2014yda} reveal a similar result, showing that the ratio of 3.5 keV emission in the core ($r<6$ arcmin) to the `confining-region' ($r>6$ arcmin) is at least a few times too large to be compatible with decaying dark matter. 

In summary, after finding no statistically significant morphological evidence for a dark matter component in Perseus or the Galactic center, we have taken a narrow window in photon energies and checked for regions of the spectrum with similar morphology to that of the 3.45-3.6 keV band.  In the Galactic center the background is highly non-thermal and we see very strong statistical evidence for correlation with known spectral lines.  At face value, the lines that most closely match the 3.5 keV morphology are those whose emissivity peaks at electron temperatures similar to that of K XVIII.  Thermal emission from such a population of $kT_e\approx2.1$~keV electrons is sub-dominant compared with the unresolved point source component. In the Perseus cluster, we see more ambiguous evidence of line correlated emission in the 3.5 keV band.  The cool thermal emission is dominant in this system and the line fluxes above 2.6 keV are very low.  Below 2.6 keV, the dense line structure also inhibits a clean correlation with continuum versus line emission.  We note, however, that either of these cases could be taken as circumstantial evidence for K XVIII, or other atomic elemental emission since, for example, a $kT_e\approx2.1$ keV electron gas maximizes the K XVIII emission and is already measured in the core of Perseus.

\section{Discussion and Conclusions}
\label{sec:conclusion}
We have examined the morphology of the X-ray emission at 3.5 keV from the Galactic center region and from the Perseus cluster of galaxies. We employed a variety of different choices to model the morphology of the continuum emission, and studied the resulting residual signal at 3.5 keV. Though it is difficult to ensure these residuals are dominated by the line signal and not mismodeling of the continuum, the azimuthal and radial distributions are strikingly different from the prediction from dark matter decay. In the case of the Galactic center, the 3.5 keV emission has a distinct quadrupolar distribution and is completely incompatible with emission from axion-like particle conversion. In the Perseus cluster 3.5 keV emission strongly overlaps with the cluster's cool clumpy core.  While the ALP scenario cannot be ruled out for Perseus, it is strongly disfavored after adding even a single additional low-energy continuum or line template.

Utilizing a sliding-window template, we demonstrated that the 3.5 keV emission most prominently correlates with the morphology of strong emission lines associated with Ar and Ca transitions in the case of the Galactic center. For Perseus, the correlation is observed in lines with a comparable peak emission temperature to K XVIII ($kT_e\approx2.16$~keV) and/or the corresponding thermal bremsstrahlung.  This is generally observed to feature a distribution overlapping with the cluster's cool core. In both the Galactic center and Perseus, we thus find strong evidence in favor of a plasma emission origin for the observed 3.5 keV line and against a dark matter interpretation.

Finally, utilizing the same binned-likelihood approach, we set the most stringent constraints to date on the lifetime of a dark matter particle decaying into a final state including a 3.5 keV monochromatic photon.  By adding a dark matter template and allowing all template normalizations to float in this process, these limits (and fits) are resilient against continuum mismodeling and robustly exclude a dark matter decay origin for the 3.5 keV line observed from clusters.

We believe that the burden of proof for a claim of discovery of any ``new'' physics must be set as high as reasonably possible. This includes pursuing with vigor Occam's razor, thus focusing with due diligence on any explanation that does not invoke unnecessary entities, and exploring with care the possible backgrounds and systematic errors. In the context of astrophysical searches for dark matter, the duck test -- ``If it looks like a duck, swims like a duck, and quacks like a duck, then it is a duck'' -- is far from sufficient proof. This work provides additional and robust evidence that the 3.5 keV line does not look, swim, nor quack like dark matter.

\section{Acknowledgments}
EC is supported by a NASA Graduate Research Fellowship under NASA NESSF Grant No. NNX13AO63H. SP is partly supported by the US Department of Energy, Contract DE-SC0010107-001.  We thank Joseph P. Conlon for feedback on an earlier version of this manuscript.
\bibliography{biblio}

\end{document}